%% file: Main.tex
\documentclass[conference]{IEEEtran}
\IEEEoverridecommandlockouts
\PassOptionsToPackage{ruled}{algorithm2e}
\usepackage[T1]{fontenc}
\usepackage[latin9]{inputenc}
\usepackage{color}
\usepackage{verbatim}
\usepackage{algorithm2e}
\usepackage{amsmath}
\usepackage{amsthm}
\usepackage{amssymb}
\usepackage{lipsum}
\makeatletter
\theoremstyle{plain}

\theoremstyle{plain}

\DeclareMathOperator{\maximize}{maximize}

\DeclareMathOperator{\diag}{diag}

\DeclareMathOperator{\tr}{Tr}
\DeclareMathOperator{\rank}{rank}

\newcommand{\trans}{^{\mathsf{T}}}

\usepackage{cite}
\usepackage{acronym}
\AtBeginDocument{\input{acro.tex}}

\usepackage{pgfplots}
\usepackage{grffile}
\pgfplotsset{compat=newest}
\usetikzlibrary{plotmarks}
\usepgfplotslibrary{patchplots}
\usepackage{amsmath}

\pgfplotsset{compat=newest}
\usetikzlibrary{plotmarks}
\usepgfplotslibrary{patchplots}

\tikzset{every node/.style={font=\small}}
\tikzset{every pin/.style={fill=white,font=\small}}
\tikzset{every pin edge/.style={<-,>=stealth,black,thick}}
\pgfplotsset{grid style={dotted,gray}}
\pgfplotsset{every axis/.style={inner sep=2pt}}
\pgfplotsset{legend style={font=\small}}
\newlength\figurewidth
\newlength\figureheight

\definecolor{mycolor1}{rgb}{1.00000,0.00000,1.00000}%

\@ifundefined{showcaptionsetup}{}{%
 \PassOptionsToPackage{caption=false}{subfig}}
\usepackage{subfig}
\makeatother

\providecommand{\lemmaname}{Lemma}
\providecommand{\theoremname}{Theorem}

\begin{document}

\title{Intelligent Omni-Surfaces (IOSs) for the MIMO Broadcast Channel
\thanks{A. Mohamed, N.~S.~Perovi\'c, M. Di Renzo are with Universit\'e Paris-Saclay, CNRS, CentraleSup\'elec,
Laboratoire des Signaux et Syst\`emes, 3 Rue Joliot-Curie, 91192
Gif-sur-Yvette, France. Email: marco.di-renzo@universite-paris-saclay.fr. This work was supported in part by the European Commission through the H2020 PAINLESS project under grant agreement number 812991, the H2020 SURFER project under grant agreement number 101030536, the H2020 ARIADNE project under grant agreement number 871464, and the H2020 RISE-6G project under grant agreement number 101017011.}
}

\author{Abdelhamed Mohamed,~Nemanja~Stefan~Perovi\'c, \IEEEmembership{Member, IEEE}, Marco~Di~Renzo,~\IEEEmembership{Fellow, IEEE}}

\maketitle	

\begin{abstract}
In this paper, we consider intelligent omni-surfaces (IOSs), which are capable of simultaneously reflecting and refracting electromagnetic waves. We focus our attention on the multiple-input multiple-output (MIMO) broadcast channel, and we introduce an algorithm for jointly optimizing the covariance matrix at the base station, the matrix of reflection and transmission coefficients at the IOS, and the amount of power that is reflected and refracted from the IOS. The distinguishable feature of this work lies in taking into account that the reflection and transmission coefficients of an IOS are tightly coupled. Simulation results are illustrated to show the convergence of the proposed algorithm and the benefits of using surfaces with simultaneous reflection and refraction capabilities.
\end{abstract}

\section{Introduction}
\bstctlcite{BSTcontrol}
Reconfigurable intelligent surfaces (RISs) have emerged as a promising approach to improve the wireless communication channel quality and to extend the network coverage \cite{di2020smart,di2019smart}. However, the vast majority of works consider surfaces that can only reflect the incident signals, which limits the coverage capabilities offered by RISs \cite{wu2019intelligent, 8741198}. To tackle this problem, the recently proposed concept of intelligent omni-surface (IOS) provides $360^\circ$ coverage thanks to surfaces that can simultaneously reflect and refract the incident signals \cite{9491943}. The reflection and refraction capabilities of the incident electromagnetic waves are controlled through the optimization of two interlinked sets of reflection and transmission coefficients. In general, in other words, it is not possible to control the reflection and transmission coefficients independently.

Recently, a few research works have analyzed the performance of IOSs. In \cite{perera2022sum}, the authors investigate the weighted sum-rate maximization under quality of service requirements and unit modulus constraints for the IOS elements, by utilizing
the successive convex approximation method. In \cite{energy-efficient}, the energy efficiency maximization problem is studied, and an optimization algorithm is proposed for jointly optimizing the transmit power and the passive beamforming at the IOS.
In \cite{niu2021simultaneous}, the weighted sum-rate of an IOS-aided multiple-input multiple-output (MIMO) system is maximized by using the alternating optimization method. The precoding matrices are obtained by the Lagrange dual method, while the reflection and transmission coefficients are obtained by the penalty concave-convex method. Interested readers can consult \cite{DBLP:journals/icl/XuLMD21}, \cite{DBLP:journals/wc/LiuMXSHPH21}, \cite{9570143}, \cite{New_Yuanwei} for further information on IOSs.

In contrast to the available works, we focus our attention on optimizing the IOS in the MIMO broadcast channel. Also, for the first time, we explicitly take into account the dependence between the reflection and transmission coefficients, based on a recently implemented IOS prototype \cite{9722826}. Specifically, the main contributions of this paper are as follows:
\begin{itemize}
\item We utilize the duality between the broadcast channel and the multiple access channel to maximize the achievable sum-rate. We formulate a joint optimization problem for the users' covariance matrices, the reflection and transmission coefficients, and the power ratio between the reflected and transmitted power. We analyze the case studies with continuous-valued and discrete-valued phase shifts for the reflection and transmission coefficients, and we assume that they are not independent of each other.
\item Due to the non-convexity of the formulated optimization problem, and the coupling between the optimization variables in the objective function, we propose an alternating optimization algorithm to solve the aforementioned problem. The optimal users' covariance matrices are obtained by applying the dual decomposition and the \ac{BCM} method, while the optimal phase shifts of the reflection and transmission coefficients of the IOS elements are formulated in a simple expression. In addition, the power ratio between the reflected and refracted power is computed iteratively by utilizing the gradient ascent method.
\item Simulation results show that the proposed algorithm converges relatively fast (i.e., within a few iterations) to a local optima. Moreover, we quantify the impact of discretizing the reflection and transmission coefficients for a two-state IOS testbed platform.
\end{itemize}

The paper is organized as follows. In Section~\ref{sec:System-Model},
we present the system model for the \mbox{IOS-aided}
MIMO broadcast channel. In Section \ref{PROBLEM FORMULATION}, we formulate the
optimization problem to maximize the achievable sum-rate. In Section
\ref{sec:Proposed-Optimization-Methods}, we describe the optimization algorithm to solve the optimization problem. In Section \ref{sec:Simulation-Results},
we provide simulation results that illustrate the achievable sum-rate. Conclusions are drawn in Section \ref{sec:Conclusion}.

\textit{Notation}: Bold upper and lower case letters denote matrices and vectors, respectively. $\mathbb{C}^{m\times n}$ denotes the
space of $m\times n$ complex matrices. $\mathbf{H}\trans$, $\mathbf{H}^H$, $|\mathbf{H}|$ and $\tr(\mathbf{H})$ 
represent the transpose, Hermitian transpose, the determinant and the trace of $\mathbf{H}$, respectively. The rank of $\mathbf{H}$ is denoted by $\rank(\mathbf{H})$ and $\lambda_{\max}(\mathbf{H})$ is the
largest singular value of $\mathbf{H}$, while $\left(\cdot\right)^{\ast}$ denotes the complex conjugate. The identity matrix is denoted by $\mathbf{I}$ and the matrix inversion operation is denoted by ${INV({\bf{X}})}$.

\section{System Model}\label{sec:System-Model}
We consider an IOS-assisted multi-user MIMO broadcast channel, where a multi-antenna \ac{BS} with $N_t$ antennas communicate with $K$ users, and each user is equipped with $N_r$ antennas. The total number of users in the reflection and transmission sides of the IOS are denoted by $KR$ and $KT$, respectively, with $K= KR + KT$. We assume that the BS and the users' antennas form a \ac{ULA}, and the inter-antenna separations are $s_t$ and $s_r$, respectively. In addition, we assume that the IOS comprises $N$ elements that can simultaneously reflect and refract the incident signals. Since the users can  be located either in the reflection or the transmission sides of the IOS, a single user receives either the signal reflected or the signal refracted from the IOS.
\begin{figure}[!t]
\centering
        \includegraphics[width=0.65\linewidth,height=0.45\linewidth]{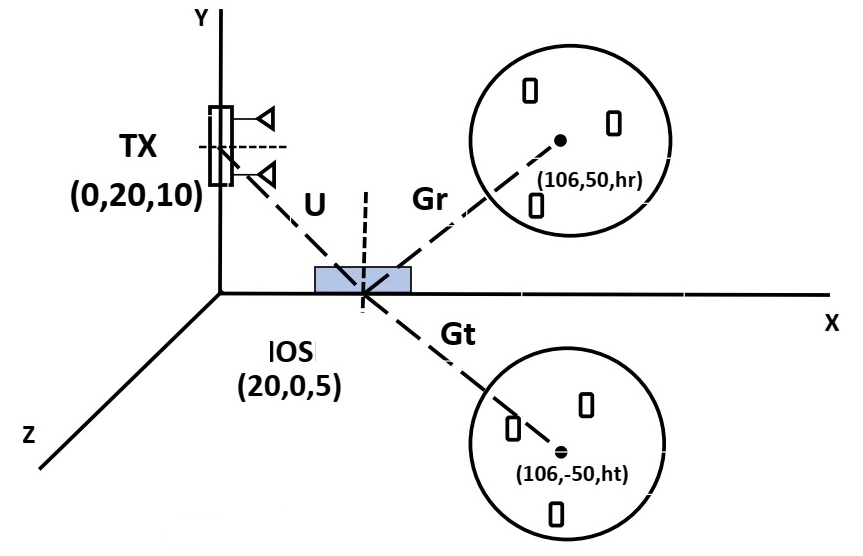} 
        \caption{ Aerial view of the considered communication system. \vspace{-0.50cm}} \label{system_model_MDR} 
\end{figure}

To serve the users located in the reflection and transmission sides simultaneously, each element of the IOS applies a complex-valued reflection and transmission coefficient. The matrix of $N$ reflection or transmission coefficients is denoted by $\mathbf{F}(\boldsymbol{\theta})=\mathrm{diag}(\boldsymbol{\theta})\in \mathbb{C}^{N\times N}$, where $\boldsymbol{\theta} =\{\boldsymbol\theta^r,\boldsymbol\theta^t\}$, and $\boldsymbol{\theta^r} = \sqrt{\rho}\left[{\Tilde{\beta}^r_{1}} , {\Tilde{\beta}^r_{2}},\dots\newline ,{\Tilde{\beta}^r_{N}} \right]^T \in \mathbb{C}^{N \times 1}$ and $\boldsymbol{\theta^t} = \sqrt{1-\rho}\left[{\Tilde{\beta}^t_{1}} , {\Tilde{\beta}^t_{2}},\dots\newline ,{\Tilde{\beta}^t_{N}} \right]^T \in \mathbb{C}^{N \times 1}$. Specifically, the parameter $\rho \in \left[0, 1 \right]$ is an optimization variable that accounts for the amount of power that the IOS directs towards the users in the reflection side ($1-\rho$ is the amount of power directed towards the transmission side). Also, $\Tilde{\beta}^{r}_n$ and $\Tilde{\beta}^{t}_n$ are the reflection and transmission coefficients of the IOS. These reflection and transmission coefficients are interlinked and are optimized in pairs for each IOS element, i.e., $\left\lbrace \Tilde{\beta}^{r}_n,\Tilde{\beta}^{t}_n\right\rbrace\in \left\lbrace({\Tilde{\beta}}^{r}_{s1},{\Tilde{\beta}}^{t}_{s1}),({\Tilde{\beta}}^{r}_{s2},{\Tilde{\beta}}^{t}_{s2}),\dots,({\Tilde{\beta}}^{r}_{sQ},{\Tilde{\beta}}^{t}_{sQ})\right\rbrace= \Psi$, where $Q$ is the number of (reflection, transmission) coefficient pairs.

Due to the presence of the IOS, the end-to-end channel for the $k_r \in [1,\dots KR]$-user located in the reflection side is: 
\begin{equation}
\begin{aligned}
&{\mathbf H_{{k_r}}} = {\mathbf{D}_{{k_r}}} + {\mathbf{G}_{{k_r}}}\mathbf F(\boldsymbol \theta^r )\mathbf{U}= {\mathbf{D}_{{k_r}}} + {\mathbf{G}_{{k_r}}}\sqrt \rho\\
&  \times \diag\left( {\tilde \beta _1^{r},\tilde \beta _2^{r},...,\tilde \beta _N^{r}} \right)\mathbf{U} = {\mathbf{D}_{{k_r}}} + \sqrt \rho  \sum\limits_{n = 1}^N {\tilde \beta _n^{r}} {\mathbf{g}_{n,{k_r}}}\mathbf{u}_n
 \end{aligned}
\end{equation}
and the end-to-end channel for the $k_t \in [1,\dots KT]$-user located in the transmission side is: 
\begin{equation}
\begin{aligned}
&{\mathbf H_{{k_t}}} = {\mathbf{D}_{{k_t}}} + {\mathbf{G}_{{k_t}}}\mathbf F(\boldsymbol \theta^t )\mathbf{U} = {\mathbf{D}_{{k_t}}} + {\mathbf{G}_{{k_t}}}\sqrt{1-\rho} \\
&\times  \diag\left( {\tilde \beta _1^{t},\tilde \beta _2^{t},...,\tilde \beta _N^{t}} \right)\mathbf{U} = {\mathbf{D}_{{k_t}}} + \sqrt{1-\rho}   \sum\limits_{n = 1}^N {\tilde \beta _n^{t}} {\mathbf{g}_{n,{k_t}}}\mathbf{u}_n
\end{aligned}
\end{equation}

In a compact form, we have:
\begin{equation}
\mathbf{H}_k(\boldsymbol{\theta})=\mathbf{D}_k + \mathbf{G}_k \mathbf{F}(\boldsymbol{\theta})\mathbf{U} 
\end{equation} 
where $\mathbf{D}_{k}\in\mathbb{C}^{N_{r}\times N_{t}}$ denotes the channel matrix between the BS and the \emph{k}-th user, $\mathbf{U}\in\mathbb{C}^{N\times N_{t}}$	denotes the channel matrix between the BS and the IOS, and $\mathbf{G}_{k}\in\mathbb{C}^{N_{r}\times N}$ denotes the channel matrix between the IOS and the \emph{k}-th user. 

To simplify the notation, we write $\mathbf{H}_{k}$ instead of $\mathbf{H}_{k}(\boldsymbol{\theta})$, where the dependence on $\boldsymbol{\theta}$ is implicit. Thus, the received signal at the $\emph{k}$-th user is written as:
\begin{equation}
\mathbf{y}_k = \mathbf{H}_k \mathbf{x}_k + \sum\nolimits_{j=1,j\neq k}^{K}\mathbf{H}_{k}\mathbf{x}_{j} + \mathbf{n}_k
\end{equation}
where $\mathbf{H}_{k}\in\mathbb{C}^{N_{r}\times N_{t}}$ is the channel matrix for the \emph{k}-th user, $\mathbf{x}_{k}\in\mathbb{C}^{N_{t}\times1}$
is the transmitted signal intended for the \emph{k}-th user, and $\mathbf{x}_{j}\in\mathbb{C}^{N_{t}\times1}$
for $j\neq k$ are the transmitted signals intended for other users, which act as interference for the detection of $\mathbf{x}_{k}$. The noise vector $\mathbf{n}_{k}\in\mathbb{C}^{N_{r}\times1}$ consists
 of \ac{iid} elements whose distribution is $\mathcal{CN}(0,N_{0})$, where $N_{0}$ is the noise variance. 

\section{Problem Formulation }\label{PROBLEM FORMULATION}
We are interested in maximizing the achievable sum-rate of the considered IOS-aided wireless communication system. To accomplish this, we exploit the fact that the achievable rate region of a Gaussian {MIMO} broadcast channel can be achieved by \ac{DPC} \cite{Weingarten:CapacityRegion:MU_MIMO:2006}.
\ac{DPC} enables us to reduce the interference in a communication system, i.e., to perfectly eliminate the interference term
$\sum_{j<k}\mathbf{H}_{k}\mathbf{x}_{j}$ for the \emph{k}-th user.
In this regard, the ordering of the users clearly matters. Let $\pi$ be an ordering of users, i.e., a permutation of the set $\{1,2,\ldots,K\}$.
Then, for this ordering, the achievable rate for the \emph{k}-th user  can be computed as \cite[Eq. (3)]{Vishwanath:duality_achievable:2003}:	\begin{equation}
R_{\pi(k)}=\log_{2}\frac{\Bigl|\mathbf{I}+\mathbf{H}_{\pi(k)}\bigl(\sum_{j\geq k}\mathbf{S}_{\pi(j)}\bigr)\mathbf{H}_{\pi(k)}^H\Bigr|}{\Bigl|\mathbf{I}+\mathbf{H}_{\pi(k)}\bigl(\sum_{j>k}\mathbf{S}_{\pi(j)}\bigr)\mathbf{H}_{\pi(k)}^H\Bigr|}
\end{equation}
where $\mathbf{S}_{k}=\mathbb{E}\bigl\{\mathbf{x}_{k}\mathbf{x}_{k}^H\bigr\}\succeq\mathbf{0}$
is the input covariance matrix of user $k$. We assume	a sum-power constraint at the BS, i.e.:
\begin{equation}
\sum\nolimits_{k=1}^{K}\tr\bigl(\mathbf{S}_{k}\bigr)\leq P
\end{equation}
where $P$ is the maximum total power at the BS. Therefore, the achievable sum-rate optimization problem for the RIS-assisted MIMO broadcast channel can be expressed as:

\begin{subequations}\label{eq:MIMO:BS:sumrate}
\begin{align}
\underset{\mathbf{S},\boldsymbol{{\Tilde{\beta}}}^l,{\rho}}{\text{max}} &~\sum_{k=1}^{K}\log_{2}\frac{\Bigl|\mathbf{I}+\mathbf{H}_{\pi(k)}\bigl(\sum_{j\geq k}\mathbf{S}_{\pi(j)}\bigr)\mathbf{H}_{\pi(k)}^H\Bigr|}{\Bigl|\mathbf{I}+\mathbf{H}_{\pi(k)}\bigl(\sum_{j>k}\mathbf{S}_{\pi(j)}\bigr)\mathbf{H}_{\pi(k)}^H\Bigr|}\\
\mathrm{s.t.} &~\sum\nolimits _{k=1}^{K}\tr\bigl(\mathbf{S}_{k}\bigr)\leq P;\mathbf{S}_{k}\succeq\mathbf{0},\forall k,\\
&\quad\left\lbrace \Tilde{\beta}^{r}_n,\Tilde{\beta}^{t}_n\right\rbrace \in \Psi;\quad \forall n\in [1,2,\dots,N],\\
&\quad 0\leq \rho \leq 1
\end{align}
\end{subequations}
where $\boldsymbol{\tilde \beta}^l =\{\boldsymbol{\tilde \beta}^r,\boldsymbol{\tilde \beta}^t\}$ and $l \in \left\lbrace r,t \right\rbrace$,.
	
To solve this problem, we exploit the duality between the MIMO broadcast channel and the multiple access channel, as recently done in \cite{perovic2021achievable}. Accordingly, we reformulate the optimization problem in (\ref{eq:MIMO:BS:sumrate}) as follows:
\begin{equation}{\small
\label{eq:MIMO:BS:sumrate:recast}
\begin{aligned}
\underset{\mathbf{\Bar{S}},\boldsymbol{{\Tilde{\beta}}}^l,{\rho}}{\max} & \quad f(\mathbf{\Bar{S}},\boldsymbol{{\Tilde{\beta}}},{\rho})=\log_{2}{\Bigl|\mathbf{I}+ \sum_{k=1}^{K}{\mathbf{H}_k^H}\mathbf{\Bar{S}}\mathbf{H}_k\Bigr|}\\
\mathrm{s.t.} & \quad\sum\nolimits _{k=1}^{K}\tr\bigl(\mathbf{\Bar{S}}_{k}\bigr)\leq P;\mathbf{\Bar{S}}_{k}\succeq\mathbf{0},\forall k,\\
&\quad\left\lbrace \Tilde{\beta}^{r}_n,\Tilde{\beta}^{t}_n\right\rbrace
\in \Psi;\quad \forall n\in [1,2,\dots,N],\\
&\quad 0\leq \rho \leq 1,
\end{aligned}}
\end{equation}
where $\mathbf{H}_k^H$ represents the dual multiple access channel corresponding to $\mathbf{H}_k$ and $\mathbf{\Bar{S}_k}$ is  the dual multiple access input covariance matrix of the \emph{k}-th user. Once the input covariance matrices $(\bar{\mathbf{S}}_{k})_{k=1}^{K}$
in the dual multiple access channel are found, the corresponding covariance matrices $(\mathbf{S}_{k})_{k=1}^{K}$
in the broadcast channel are obtained from \cite[Eq. (11)]{Vishwanath:duality_achievable:2003}, as:
\begin{equation}
\mathbf{S}_{k}=\mathbf{B}_{k}^{-1/2}\mathbf{F}_{k}\mathbf{G}_{k}^H\mathbf{A}_{k}^{1/2}\bar{\mathbf{S}}_{k}\mathbf{A}_{k}^{1/2}\mathbf{G}_{k}\mathbf{F}_{k}^H\mathbf{B}_{k}^{-1/2}\label{eq:Cov_mat_rel}
\end{equation}
where $\mathbf{A}_{k}=\mathbf{I}+\mathbf{H}_{k}(\sum_{i=1}^{k-1}\mathbf{S}_{i})\mathbf{H}_{k}^H$, $\mathbf{B}_{k}=\mathbf{I}+\sum_{i=k+1}^{K}\mathbf{H}_{i}^H\bar{\mathbf{S}}_{i}\mathbf{H}_{i}$,
and  $\mathbf{F}_{k}\boldsymbol{\Lambda}_{k}\mathbf{G}_{k}^H$ is the singular value decomposition of $\mathbf{B}_{k}^{-1/2}\mathbf{H}_{k}^H\mathbf{A}_{k}^{-1/2}$. 

\section{Proposed Optimization Method\label{sec:Proposed-Optimization-Methods}}
To solve the formulated problem, we propose an alternating optimization method. The users' covariance matrices are optimized by exploiting the dual decomposition method in \cite{perovic2021achievable}. The reflection and transmission coefficients of the IOS are obtained by generalizing the method in \cite{zhang2019capacity}, which is applicable to reflecting surfaces, under the assumption of continuous-valued coefficients. The corresponding discrete-valued reflection and transmission coefficients are obtained by projecting the obtained solutions onto the feasible set of possible discrete values. Moreover, we present a gradient-based method for optimizing the power ratio. 

\subsection{Covariance Matrix Optimization \label{subsec:AO-Cov-Mat-PGM}}
For given values of the power ratio and the reflection and transmission coefficients, the achievable sum-rate optimization
problem in \eqref{eq:MIMO:BS:sumrate:recast} is simplified as follows: \vspace{-0.25cm} \begin{subequations}\label{eq:MIMO:MAC:fixtheta}{
\begin{align}
\underset{\bar{\mathbf{S}}}{\max} & \quad\log_{2}\Bigl|\mathbf{I}+\sum\limits _{k=1}^{K}\mathbf{H}_{k}^H\bar{\mathbf{S}}_{k}\mathbf{H}_{k}\Bigr|\\
\mathrm{s.t.} & \quad\bar{\mathbf{S}}\in\mathcal{S}\label{eq:MAC:SPC}
\end{align}
}\end{subequations}
where  $\bar{\mathbf{S}}\triangleq(\bar{\mathbf{S}}_{k})_{k=1}^{K}$ for 
$\mathcal{S}=\{\bar{\mathbf{S}}\ |\ \sum\nolimits _{k=1}^{K}\tr\bigl(\bar{\mathbf{S}}_{k}\bigr)\leq P;\bar{\mathbf{S}}_{k}\succeq\mathbf{0}\thinspace\thinspace\forall k\}$. As described in \cite{perovic2021achievable, perovic2021maximum}, the optimization problem in (10) is solved by using the dual decomposition Lagrangian and the accelerated block coordinate maximization methods.

\subsection{IOS Optimization}
For given $\left\{ {{{\bar S}_k}} \right\}_{k = 1}^K$ and $\rho$, the optimization
problem in \eqref{eq:MIMO:BS:sumrate:recast} with respect to $\Tilde{\beta}_n^{l}$, $l \in \left\lbrace r,t \right\rbrace$, can be written as follows:
\begin{equation}{\small
\begin{aligned}
\underset{{{\Tilde{\beta}_n^l}}}{\max} & \quad f({{\Tilde{\beta}_n^l}})=\log_{2}{\Bigl|\mathbf{I}+ \sum_{k=1}^{K}{\mathbf{H}_k^H}\mathbf{\Bar{S}}\mathbf{H}_k\Bigr|}\\
\mathrm{s.t.} &\quad\left\lbrace \Tilde{\beta}^{r}_n,\Tilde{\beta}^{t}_n\right\rbrace \in \Psi;\quad \forall n\in [1,2,\dots,N] \label{eq:MIMO:BS:sumrate:theta}
\end{aligned}}
\end{equation}

To tackle this problem, we first optimize both $\Tilde{\beta}^{r}_n$ and $\Tilde{\beta}^{t}$ by assuming that they are continuous-valued and independent coefficients, and we then project the obtained solutions onto the set of feasible discrete phase shifts, by taking into account that the reflection and transmission coefficients are interlinked. 

For ease of writing, we define the variables
${{\bf{G}}_{{k}}} = [{\bf{g}}_1,  \ldots ,{\bf{g}}_n]$, $ {\bf{U}} = {[{\bf{u}}_{1}^T,...,{\bf{u}}_n^T]^T}$, and

\begin{equation}
{{{\bf{\bar D}}}_k} = \left\{ \begin{array}{l}
{\bf{D}}_i^r~~\forall i \in \left\{ {1,...,{KR}} \right\}\\
{\bf{D}}_j^t~~\forall j \in \left\{ {1,...,{KT}} \right\}
\end{array} \right.
\end{equation}
\begin{equation}
{{{\bf{\bar g}}}}_{n,k} = \left\{ \begin{array}{l}
{\bf{g}}_{n,i}^r~~ \forall i \in \left\{ {1,...,{KR}} \right\}\\
{\bf{g}}_{n,j}^t~~\forall j \in \left\{ {1,...,{KT}} \right\}
\end{array} \right.
\end{equation}

Under the assumption that $\Tilde{\beta}^{r}_n$ and $\Tilde{\beta}^{t}_n$ can be optimized independently (this is removed next by projecting on the feasible set),
the objective function can be rewritten as follows: 
\begin{equation}{\small\label{eq:MIMO:BS:sumrate:theta:recast}
f({{\Tilde{\beta}_n^l}})=
 {\log _2}\left| \begin{array}{l}
{\bf{I}} + \left( {\begin{array}{*{20}{l}}
{\sum\limits_{k = 1}^{KR} {\left( {{{{\bf{\bar D}}}_k} + \sum\limits_{n = 1}^N {\sqrt \rho  \tilde \beta _n^r} {{{\bf{\bar g}}}_{n,k}}{{\bf{u}}_n}} \right)^H} }\\
{ \times {{\overline {\bf{S}} }_k}{{\left( {{{{\bf{\bar D}}}_k} + \sum\limits_{n = 1}^N {\sqrt \rho  \tilde \beta _n^r} {{{\bf{\bar g}}}_{n,k}}{{\bf{u}}_n}} \right)}}}
\end{array}} \right)\\
 + \left( {\begin{array}{*{20}{l}}
{\sum\limits_{k = 1}^{KT} {\left( {{{{\bf{\bar D}}}_k} + \sum\limits_{n = 1}^N {\sqrt {1 - \rho } \tilde \beta _n^t} {{{\bf{\bar g}}}_{n,k}}{{\bf{u}}_n}} \right)^H} }\\
{ \times {{\overline {\bf{S}} }_k}{{\left( {{{{\bf{\bar D}}}_k} + \sum\limits_{n = 1}^N {\sqrt {1 - \rho } \tilde \beta _n^t} {{{\bf{\bar g}}}_{n,k}}{{\bf{u}}_n}} \right)}}}
\end{array}} \right)
\end{array} \right|}
\end{equation}

Thus, for optimizing $\tilde \beta _n^r$, we have:
\begin{equation}
f(\tilde \beta _n^r) = {\log _2}\left| {{\bf{A}}_n^r + \tilde \beta _n^r{\bf{B}}_n^r + \tilde \beta _n^{r * }{\bf{B}}_n^{rH}} \right|
\end{equation}
and, for optimizing $\tilde \beta _n^t$, we have:
\begin{equation}
f(\tilde \beta _n^t) = {\log _2}\left| {{\bf{A}}_n^t + \tilde \beta _n^t{\bf{B}}_n^t +  \tilde \beta _n^{t * }{\bf{B}}_n^{tH}} \right|\end{equation}
where:
\begin{equation}{\small
{\bf{B}}_n^r = \sqrt {\rho } \sum\limits_{k = 1}^{KR} {( {{{{\bf{\bar D}}}^H}_k + \sum\limits_{l = 1,n \ne l}^N {\sqrt \rho  \tilde \beta _l^{r*}} {\bf{u}}_l^H{{{\bf{\bar g}}}^H}_{l,k}} ){{\overline {\bf{S}} }_k}{{{\bf{\bar g}}}_{n,k}}{{\bf{u}}_n}} }
\end{equation}
\begin{equation}{\small
{\bf{B}}_n^t =  \sum\limits_{k = 1}^{KT} {(\sqrt {1 - \rho } {{{{\bf{\bar D}}}^H}_k + \sum\limits_{n \ne l}^N {({1 - \rho) } \tilde \beta _l^{t*}} {\bf{u}}_l^H{{{\bf{\bar g}}}^H}_{l,k}} ){{\overline {\bf{S}} }_k}{{{\bf{\bar g}}}_{n,k}}{{\bf{u}}_n}} }                     
\end{equation}
\begin{equation}{\small
\begin{array}{l}
{\bf{A}}_n^r = {\bf{I}} + \sum\limits_{k = 1}^{KT} {( {{{{\bf{\bar D}}}_k} + \sum\limits_{n = 1}^N {\sqrt {1 - \rho } \tilde \beta _n^t} {{{\bf{\bar g}}}_{n,k}}{{\bf{u}}_n}})^H} \\
 \times {\overline {\bf{S}} _k}{( {{{{\bf{\bar D}}}_k} + \sum\limits_{n = 1}^N {\sqrt {1 - \rho } \tilde \beta _n^t} {{{\bf{\bar g}}}_{n,k}}{{\bf{u}}_n}} )} + \sum\limits_{k = 1}^{KR} \rho  {\bf{u}}_n^H{{{\bf{\bar g}}}^H}_{n,k}{\overline {\bf{S}} _k}{{{\bf{\bar g}}}_{n,k}}{{\bf{u}}_n}\\
 + \sum\limits_{k = 1}^{KR} { ( {{{{\bf{\bar D}}}^H}_k + \sum\limits_{l = 1,n \ne l}^N {\sqrt \rho  \tilde \beta _l^{r*}{\bf{u}}_l^H{{{\bf{\bar g}}}^H}_{l,k}} })} \\
 \times {\overline {\bf{S}} _k} ( {{{{\bf{\bar D}}}_k} + \sum\limits_{l = 1,n \ne l}^N {\sqrt \rho  \tilde \beta _l^r{{{\bf{\bar g}}}_{l,k}}{{\bf{u}}_l}} } )
\end{array}}
\end{equation}
\begin{equation}{\small
\begin{array}{*{20}{l}}
\begin{array}{l}
{\bf{A}}_n^t = {\bf{I}} + \sum\limits_{k = 1}^{KR} {( {{{{\bf{\bar D}}}_k} + \sum\limits_{n = 1}^N {\sqrt \rho  \tilde \beta _n^r} {{{\bf{\bar g}}}_{n,k}}{{\bf{u}}_n}})^H} \\
 \times {\overline {\bf{S}} _k}{( {{{{\bf{\bar D}}}_k} + \sum\limits_{n = 1}^N {\sqrt \rho  \tilde \beta _n^r} {{{\bf{\bar g}}}_{n,k}}{{\bf{u}}_n}} )} + \sum\limits_{k = 1}^{KT} {( {1 - \rho } )} {\bf{u}}_n^H{{{\bf{\bar g}}}^H}_{n,k}{\overline {\bf{S}} _k}{{{\bf{\bar g}}}_{n,k}}{{\bf{u}}_n}\\
 + \sum\limits_{k = 1}^{KT} {  ( {{{{\bf{\bar D}}}^H}_k + \sum\limits_{l = 1,n \ne l}^N {\sqrt {1 - \rho } \tilde \beta _l^{t*}{\bf{u}}_l^H{{{\bf{\bar g}}}^H}_{l,k}} } )} 
\end{array}\\
{ \times {{\overline {\bf{S}} }_k} ( {{{{\bf{\bar D}}}_k} + \sum\limits_{l = 1,n \ne l}^N {\sqrt {1 - \rho } \tilde \beta _l^t{{{\bf{\bar g}}}_{l,k}}{{\bf{u}}_l}}})}
\end{array}}
\end{equation}

\begin{algorithm}[t]
{\small\caption{AO algorithm to solve  \eqref{eq:MIMO:BS:sumrate:recast}\label{Overall_algorithm}}

\SetAlgoNoLine
\DontPrintSemicolon
\LinesNumbered

\KwIn{$\bar{\mathbf{S}}^{(0)}$, $\left\{\Tilde{\beta}^{r(0)}, \Tilde{\beta}^{t(0)}\right\}$ and  $\rho^{(0)}$ with feasible values}

Set $i\leftarrow 0$

\Repeat{a stopping criterion is met}{

Compute $(\bar{\mathbf{S}}_{k}^{(i+1)})_{k=1}^{K}$ according to \textbf{Algorithm 2} in \cite{perovic2021achievable}
\\
\For{$n=1,2,\ldots,N_{\mathrm{ris}}$}{

$ \Tilde{\beta}_n^{*(i+1)}=\exp(-j\arg(\sigma_{n}))$ using \eqref{optimal solution phase shift}\\
Apply the phase shift projection procedure, if required, based on \eqref{proj:phase:shift}}

$\rho^{{(i+1)}}=\mathbb{P}_D\left({\rho}^{(i)}+{\upsilon^{(i)}}\nabla_\rho f(\rho)|_{\rho={\rho^{(i)}}}\right)$

$ i\leftarrow i+1 $;}
\KwOut{${\mathbf{S}}^{\bullet}={\mathbf{S}}^{(i)}$,~$\rho^{\bullet}=\rho^{(i)}$ and $\left\{ {{{{\bf{\tilde \beta }}}^{r \bullet(i) }},{{{\bf{\tilde \beta }}}^{t \bullet (i) }}} \right\}$}}
\end{algorithm}

The optimal solution of the optimization problem in \eqref{eq:MIMO:BS:sumrate:theta:recast} is then given by \cite{perovic2021achievable}:
\begin{subequations}{\small
\begin{align}
&\tilde \beta _n^{*r} = \exp ( - j\arg (\sigma _n^r))\\
&\tilde \beta _n^{*t} = \exp ( - j\arg (\sigma _n^t))
\end{align}\label{optimal solution phase shift}}
\end{subequations} 
 where $\sigma _n^r$ and $ \sigma _n^t$ are the only non-zero eigenvalues of ${\left( {{\bf{A}}_n^r} \right)^{ - 1}}{\bf{B}}_n^r$ and ${\left( {{\bf{A}}_n^t} \right)^{ - 1}}{\bf{B}}_n^t$, respectively. 

The obtained continuous-valued solution of the reflection and transmission coefficients is denoted by $\left\{\Tilde{\beta}_n^{*r}, \Tilde{\beta}_n^{*t}\right\}=SC$. The corresponding discrete-valued solution in the considered feasible set is obtained by  projecting $SC$ on the discrete set $\Psi$. As a case study, we consider the IOS prototype in \cite{9722826}, where each IOS element can be configured in two states, where each state is identified by a pair of reflection and transmission coefficients, i.e.,  $\left\{({{{\bf{\tilde \beta }}}^{r  }_{S1}},{{{\bf{\tilde \beta }}}^{t}_{S1}}),({{{\bf{\tilde \beta }}}^{r}_{S2}},{{{\bf{\tilde \beta }}}^{t }_{S2}})\right\}=\left\{ S1,S2 \right\} =\Psi $.

In this case, the projection can be formulated as follows:
\begin{equation}{
\left\{ {{{{\bf{\tilde \beta }}}^{r \bullet }},{{{\bf{\tilde \beta }}}^{t \bullet }}} \right\} = \left\{ \begin{array}{l}
\left( {\beta _{S1}^r,\beta _{S1}^t} \right),{\left\| {SC-S1} \right\|_2} < {\left\| {SC-S2} \right\|_2}\\
\left( {\beta _{S2}^r,\beta _{S2}^t} \right),\quad \text{otherwise}
\end{array} \right.
\label{proj:phase:shift}}
\end{equation} where $\left\{ {{{{\bf{\tilde \beta }}}^{r \bullet }},{{{\bf{\tilde \beta }}}^{t \bullet }}} \right\}$ denotes the reflection and transmission coefficients that minimize the distance between the solution of the continuous-valued case and the feasible discrete set.

\subsection{Power Ratio Optimization}
For given values of $\left\{ {{{\bar S}_k^*}} \right\}_{k = 1}^K$ and $\left\{ {{{{\bf{\tilde \beta }}}^{r \bullet }},{{{\bf{\tilde \beta }}}^{t \bullet }}} \right\}$, the optimization problem in \eqref{eq:MIMO:BS:sumrate:recast} 
can be explicitly rewritten as: \begin{subequations}{\small\label{eq:MIMO:MAC:fixcovar&beta}
\begin{align}
\underset{\rho}{\max} & \quad f(\rho)=\log_{2}\Bigl|\mathbf{I}+\sum\limits _{k=1}^{K}\mathbf{H}_{k}^H\bar{\mathbf{S}}_{k}\mathbf{H}_{k}\Bigl|\label{eq:OP:problem:rho}\\
\mathrm{s.t.} & \quad 0\leq \rho \leq 1
\end{align}
}\end{subequations}  

The objective function $f(\rho)$ can be rewritten as:
\begin{equation}{\small
\begin{aligned}
&f(\rho ) = {\log_2} \left| {{\bf{I}} + \sum\limits_{k = 1}^K {{{\bf{H}}_k^H}{{\overline {\bf{S}} }_k}{\bf{H}}_k} } \right|\end{aligned}}\end{equation}
 \begin{equation}{\small
 = \log_2  \left| \begin{array}{c}
\begin{array}{*{20}{l}}
\begin{array}{l}
{\bf{I}} + 2\sqrt \rho  \sum\limits_{k = 1}^{KR} {\sum\limits_{n = 1}^N {{{\tilde \beta }_n}^{r*} {{\bf{u}}_n^H}{{{\bf{\bar g}}}_{n,k}^H}{{\overline {\bf{S}} }_k}{{{\bf{\bar D}}}_k}}} \\
 + 2\sqrt {1 - \rho } \sum\limits_{k = 1}^{KT} {\sum\limits_{n = 1}^N {{{\tilde \beta }_n}^{t*}}{{\bf{u}}_n^H} {{{\bf{\bar g}}}_{n,k}^H}{{\overline {\bf{S}} }_k}{{{\bf{\bar D}}}_k}}
\end{array}\\
+{\sum\limits_{k = 1}^{KR} {\left( {\begin{array}{*{20}{l}}
{\sum\limits_{n = 1}^N {\sum\limits_{n = 1}^N {\rho {{\tilde \beta }_n^{r*}}\tilde \beta _n^{r}{{\bf{u}}_n^H}} } {{{\bf{\bar g}}}_{n,k}^H}{{\overline {\bf{S}} }_k}{\bf{\bar g}}_{n,k}{\bf{u}}_n}\\
{ + {{{\bf{\bar D}}}_k^H}{{\overline {\bf{S}} }_k}{{{\bf{\bar D}}}_k}}
\end{array}} \right)} }
\end{array}\\
 + \sum\limits_{k = 1}^{KT} {\left( {\begin{array}{*{20}{l}}
{\sum\limits_{n = 1}^N {\sum\limits_{n = 1}^N {(1 - \rho ){{\tilde \beta }_n^{t*}}\tilde \beta _n^{t} } {\bf{u}}_n^H{\bf{\bar g}}_{n,k}^H{{\overline {\bf{S}} }_k}}{{{\bf{\bar g}}}_{n,k}}}{{\bf{u}}_n}\\
{ + {{{\bf{\bar D}}}_k^H}{{\overline {\bf{S}} }_k}{{{\bf{\bar D}}}_k}}
\end{array}} \right)} 
\end{array} \right|}
\end{equation}

In order to find the optimal value of the power ratio $\rho$, we solve the equation $f^{'}(\rho)=0$, where $f^{'}(\rho)$ is the first-order derivative of $f(\rho)$. 
For ease of notation, let us define:
\begin{equation}{\small
{{\bf{X}}_{\bf{1}}} = \sum\nolimits_{k = 1}^{KR} {\sum\nolimits_{n = 1}^N {{{\tilde \beta }_n}^{r*} {{\bf{u}}_n^H}{{{\bf{\bar g}}}_{n,k}^H}{{\overline {\bf{S}} }_k}{{{\bf{\bar D}}}_k}}}}\end{equation}
\begin{equation}{\small
{{\bf{X}}_{\bf{2}}} =  \sum\nolimits_{k = 1}^{KT} {\sum\nolimits_{n = 1}^N {{{\tilde \beta }_n}^{t*}}{{\bf{u}}_n^H} {{{\bf{\bar g}}}_{n,k}^H}{{\overline {\bf{S}} }_k}{{{\bf{\bar D}}}_k}}} \end{equation}
\begin{equation}{\small
{{\bf{Y}}_{\bf{1}}} = \sum\nolimits_{k = 1}^{KR} {\sum\nolimits_{n = 1}^N {\sum\nolimits_{n = 1}^N { {{\tilde \beta }_n^{r*}}\tilde \beta _n^{r}{{\bf{u}}_n^H}} } {{{\bf{\bar g}}}_{n,k}^H}{{\overline {\bf{S}} }_k}{\bf{\bar g}}_{n,k}{\bf{u}}_n} }\end{equation}
\begin{equation}{\small
{{\bf{Y}}_2} = \sum\nolimits_{k = 1}^{KT} {\sum\nolimits_{n = 1}^N {\sum\nolimits_{n = 1}^N {{{\tilde \beta }_n^{t*}}\tilde \beta _n^{t} } {\bf{u}}_n^H{\bf{\bar g}}_{n,k}^H{{\overline {\bf{S}} }_k}}{{{\bf{\bar g}}}_{n,k}}}{{\bf{u}}_n} } \end{equation}
\begin{equation}{\small
\begin{aligned}
{\bf{\bar X}} &={{\bf{I}} + \sum\nolimits_{i = 1}^{{KR}} {{{\bf{H}}_i}{{\overline {\bf{S}} }_i}{\bf{H}}_i^H}  + \sum\nolimits_{j = 1}^{{KT}} {{{\bf{H}}_j}{{\overline {\bf{S}} }_j}{\bf{H}}_j^H} }
\end{aligned}}
\end{equation}

\begin{figure*}[!t]
    \centering
    \begin{minipage}{0.40\linewidth}
        \centering
        \includegraphics[width=0.8\linewidth,height=0.60\linewidth]{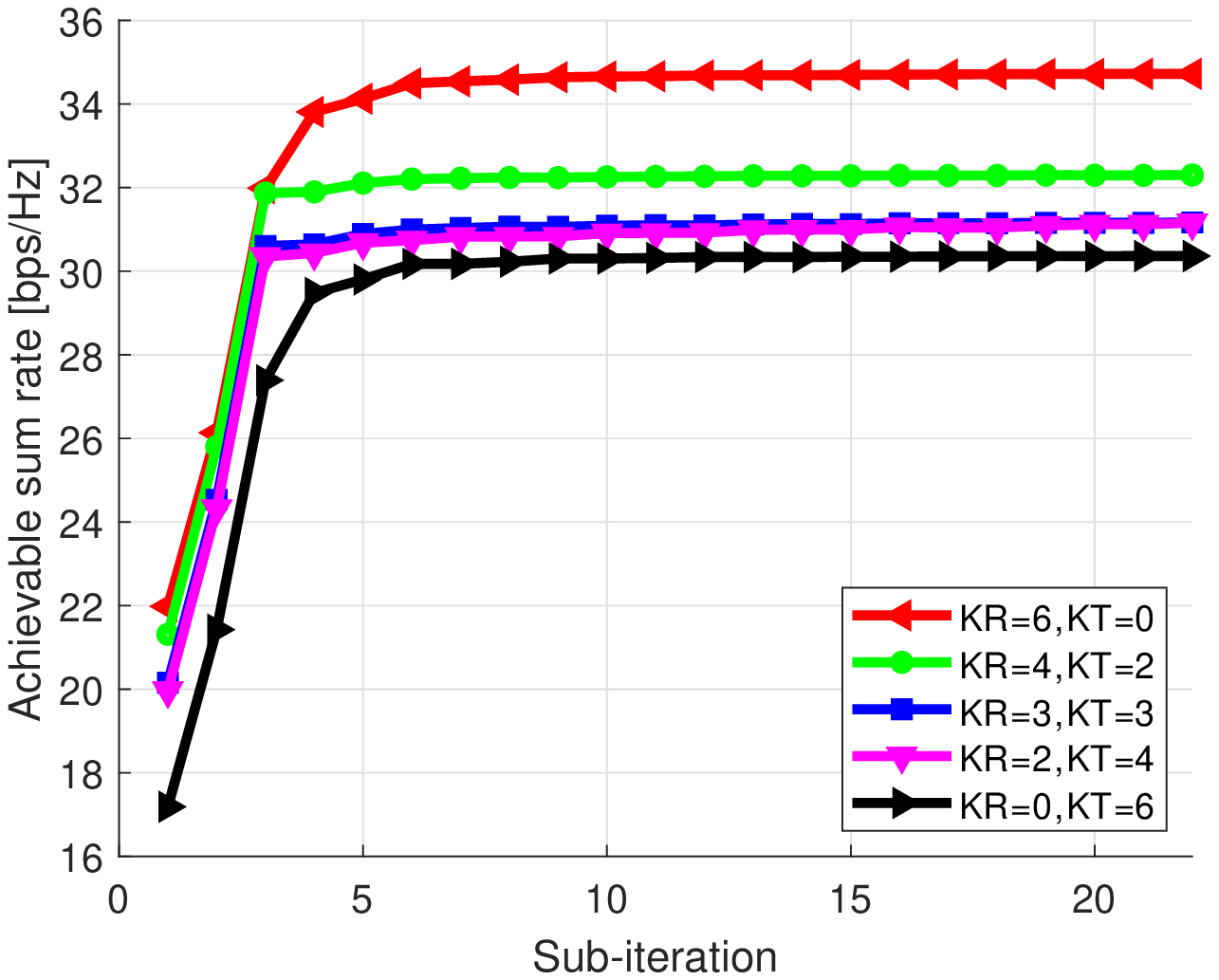} 
        \caption{Continuous-valued coefficients.} \label{Continuous Scenario}
    \end{minipage}  \hspace*{\fill}
    \begin{minipage}{0.40\linewidth}
        \centering
        \includegraphics[width=0.80\linewidth,height=0.60\linewidth]{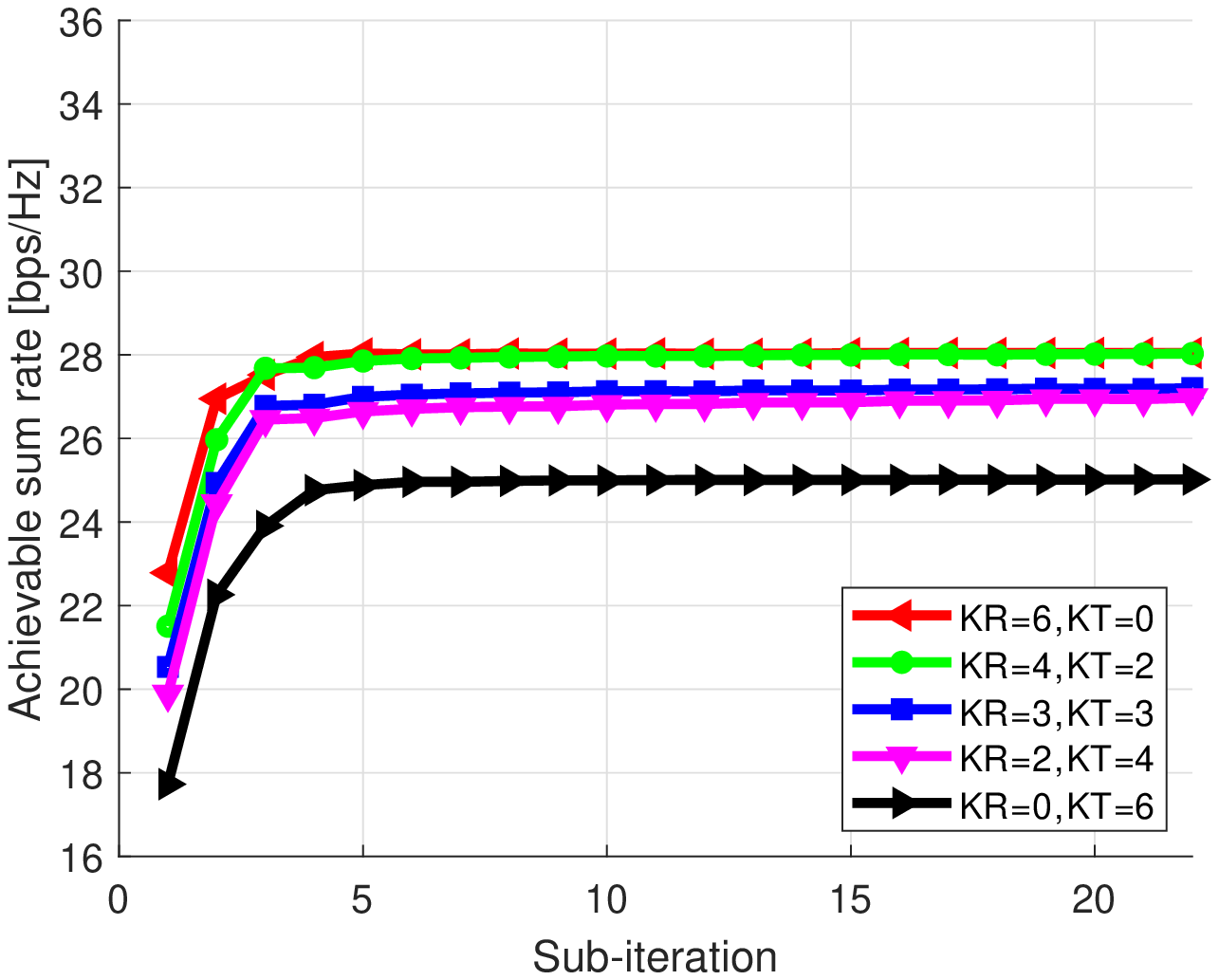} 
        \caption{Discrete-valued coefficients.} \label{Discrete Scenario}
    \end{minipage} \hspace*{\fill}
\end{figure*}

The first-order and second-order derivatives, $f^{'}(\rho)$ and $f^{''}(\rho)$, respectively, can be formulated as:
\begin{equation}{
f'(\rho ) =\nabla_\rho f(\rho)= \tr\left( {{{{\bf{\bar X}}}^{ - 1}}\left( \begin{array}{l}
\left( {\frac{{{{\bf{X}}_1}}}{{\sqrt \rho  }} - \frac{{{{\bf{X}}_2}}}{{\sqrt {1 - \rho } }}} \right)\\
 + \left( {{{\bf{Y}}_1} - {{\bf{Y}}_2}} \right)
\end{array} \right)} \right)}\label{1st derivative}
\end{equation}
\begin{equation}{
\begin{array}{l}
f''(\rho ) =  - \frac{1}{{\ln 2}}{\left( {INV({\bf{\bar X}})} \right)^2}\\
 \times \left( \begin{array}{l}
0.5 {\left( {\sqrt[{ - 3/2}]{\rho }{{\bf{X}}_{\bf{1}}} + \sqrt[{ - 3/2}]{{1 - \rho }}{{\bf{X}}_{\bf{2}}}} \right) {{\bf{\bar X}}}}\\
 + {\left( {\left( {\frac{{{{\bf{X}}_1}}}{{\sqrt \rho  }} - \frac{{{{\bf{X}}_2}}}{{\sqrt {1 - \rho } }}} \right) + \left( {{{\bf{Y}}_1} - {{\bf{Y}}_2}} \right)} \right)^2}
\end{array} \right)
\end{array}}\label{2nd derivative}
\end{equation}
where $\nabla_\rho f(\rho)$ is the gradient of $f(\rho )$ with respect to $\rho$.

The objective function in \eqref{eq:OP:problem:rho} is a concave function with respect to $\rho$, since the second derivative in \eqref{2nd derivative} is no greater than zero. Therefore, the optimization problem in \eqref{eq:OP:problem:rho} has a single optimum value. However, it is hard to find a closed-form solution by solving $f^{'}(\rho)=0$. Thus, we utilize the projected gradient (PG) method. Specifically, $\rho$ in the $(i+1)$-th iteration is updated as follows: 
\begin{equation}
\rho^{i+1}=\mathbb{P}_D\left({\rho}^i+{\upsilon^i}\nabla_\rho f(\rho)|_{\rho={\rho^i}}\right)\label{update_rho}
\end{equation}
where $\upsilon^i$ in \eqref{update_rho} is the step size that is updated by using the backtracking line search method in \cite{perovic2021achievable}, and
the projection operator $\mathbb{P}_D$ is defined as:
\begin{equation}{\small
 \hat{\rho}= \mathbb{P}_D(\rho) = \left\{ {\begin{array}{*{20}{l}}
 	{\rho^{min}}\qquad &{\rho^ \bullet  < \rho ^{min}} \\ 
 	{\rho ^ \bullet}\qquad&{\rho ^{min} \leqslant \rho ^ \bullet  \leqslant \rho ^{max}} \\ 
 	{\rho ^{max}}\qquad&{ {\rho } > \rho ^{max}}
 	\end{array}} \right.\label{ps_updating_criteria}}
  \end{equation} 
where $\rho^\bullet$ and $\hat\rho$  are the solutions obtained before and after implementing the projection, and $\rho^{min}$ and $\rho^{max}$ are the minimum and maximum values for $\rho$, respectively.
\section{Simulation Results\label{sec:Simulation-Results}}
In this section, we evaluate the achievable sum-rate of the proposed
algorithms with the aid of Monte Carlo simulations. Specifically, we compare the sum-rates under the assumption that the reflection and transmission coefficients are continuous-valued and independent values, and under the assumption that they are interlinked and belong to a discrete set.

The simulation setup is the following: the carrier frequency is $f=2\,\mathrm{GHz}$
(the wavelength is $\lambda=15\,\mathrm{cm}$), $s_{t}=s_{r}=s_{\mathrm{ris}}=\lambda/2=7.5\,\mathrm{cm}$, the network topology is given in Fig. 1, $N_{t}=8$,
$P=1\,\mathrm{W}$, and $N_{0}=-110\thinspace\mathrm{dB}$. The IOS
consists of $N_{\mathrm{ris}}=15\times15=225$ elements placed in a \mbox{$15\times15$}
square formation. The users are equipped with $N_{r}=2$
antennas and are randomly distributed within the disks shown in Fig. 1. The results are averaged over 100 independent channel realizations. 

The achievable sum-rate is reported in Fig. \ref{Continuous Scenario} and Fig. \ref{Discrete Scenario}. The proposed algorithm converges in a few iterations. By closing analyzing the sum-rate of the users located in the reflection and transmission sides of the IOS, we observe that the sum-rate loss in Fig. \ref{Discrete Scenario} as compared to Fig. \ref{Continuous Scenario} is due to the non-unitary value of the reflection and transmission coefficients for the considered feasible set in \cite[Table 1]{9722826}.

\section{Conclusion\label{sec:Conclusion}}
In this paper, we have proposed optimization algorithms for maximizing the sum-rate in IOS-aided MIMO broadcast channels. We have optimized the covariance matrices at the transmitter, the reflection and transmission coefficients of the IOS, and the ratio between the reflected and refracted power. The simulations results demonstrate that the presented algorithm provides a rapid convergence rate in a few iterations.

\bibliographystyle{IEEEtran}
\bibliography{IEEEabrv,Main,IEEEexample}
	
\end{document}

%% file: acro.tex
\acrodef{AO}{alternating optimization}
\acrodef{APGM}{alternating projected gradient method}
\acrodef{APM}{accelerated proximal gradient method}
\acrodef{ASP}{antenna separation product}
\acrodef{AWGN}{additive white Gaussian noise}
\acrodef{BC}{broadcast channel}
\acrodef{BCM}{block coordinate maximization}
\acrodef{BEP}{bit error probability}
\acrodef{BER}{bit error rate}
\acrodef{BF-MIMO}[BF\mbox{-}MIMO]{beamforming MIMO}
\acrodef{BF}{beamforming}
\acrodef{BS}{base station}
\acrodef{bpcu}{bits per channel use}
\acrodef{CP}{cyclic prefix}
\acrodef{CSI}{channel state information}
\acrodef{CSIR}{channel state information at RX}
\acrodef{SSK}{space shift keying}
\acrodef{CSIT}{channel state information at TX}
\acrodef{DCMC}{discrete\mbox{-}input continuous\mbox{-}output memoryless channel}
\acrodef{DFT}{discrete Fourier transform}
\acrodef{DL-TR-GSM}{dual-layered transmit-receive \acl{GSM}}
\acrodef{DLT}{dual-layered transmission}
\acrodef{DPC}{dirty paper coding}
\acrodef{EGC}{equal gain combining}
\acrodef{EM}{electromagnetic}
\acrodef{EVD}{eigenvalue decomposition}
\acrodef{FSPL}{free space path loss}
\acrodef{FFT}{fast Fourier transform}
\acrodef{FDE}{frequency domain equalization}
\acrodef{GRSM}{generalized \acl{RSM}}
\acrodef{GSM}{generalized \acl{SM}}
\acrodef{IFFT}{invserse fast Fourier transform}
\acrodef{ICI}{inter-channel interference}
\acrodef{iid}[i.i.d.]{independent and identically distributed}
\acrodef{IOS}{intelligent omni\mbox{-}surface}
\acrodef{IQ}{in\mbox{-}phase and quadrature}
\acrodef{ISI}{intersymbol interference}
\acrodef{ISI-free}[ISI\mbox{-}free]{intersymbol interference free}
\acrodef{LIS}{large intelligent surface}
\acrodef{LOS}{line\mbox{-}of\mbox{-}sight}
\acrodef{KKT}{Karush\mbox{-}Kuhn\mbox{-}Tucker} 
\acrodef{MAC}{multiple-access channel}
\acrodef{mmWave}{millimeter-wave}
\acrodef{MIMO}{multiple\mbox{-}input multiple\mbox{-}output}
\acrodef{MISO}{multiple\mbox{-}input single\mbox{-}output}
\acrodef{ML}{maximum likelihood}
\acrodef{MRC}{maximal ratio combining}
\acrodef{MMSE}{minimum mean square error}
\acrodef{MU-TR-GSM}{multiuser transmit-receive  \acl{GSM} }
\acrodef{NCSIT}{no channel state information at TX}
\acrodef{NLOS}{non\mbox{-}\acs{LOS}} 
\acrodef{NOMA}{non-orthogonal multiple access}
\acrodef{OFDM}{orthogonal frequency division multiplexing}
\acrodef{OFDMA}{orthogonal frequency division multiple access}
\acrodef{PA}{power amplifier}
\acrodef{PAE}{power added efficiency}
\acrodef{PAPR}{peak\mbox{-}to\mbox{-}average power ratio}
\acrodef{PDF}{probability density function}
\acrodef{PEP}{pairwise error probability}
\acrodefplural{PEP}{pairwise error probabilities}
\acrodef{PGM}{projected gradient method}
\acrodef{PMP}{probability mass function}
\acrodef{PSM}{precoding-aided spatial modulation}
\acrodef{QSM}{quadrature spatial modulation}
\acrodef{RC}{reorganization computation}
\acrodef{RIS}{reconfigurable intelligent surface}
\acrodef{RSM}{receive spatial modulation}
\acrodef{RX}{receiver}
\acrodef{SEP}{symbol error probability}
\acrodef{SER}{symbol error rate}
\acrodef{SIC}{successive interference cancellation}
\acrodef{SINR}{signal-to-interference-plus-noise ratio}
\acrodef{SISO}{single-input single-output}
\acrodef{SM}{spatial modulation}
\acrodef{SMX-MIMO}[SMX\mbox{-}MIMO]{spatial multiplexing MIMO}
\acrodef{SMX}{spatial multiplexing}
\acrodef{SNR}{signal-to-noise ratio}
\acrodef{SC}{single carrier}
\acrodef{SVD}{singular value decomposition}
\acrodef{SPST}{single pole single-throw}
\acrodef{SU}{secondary user}
\acrodef{TDE}{time domain equalization}
\acrodef{TX}{transmitter}
\acrodef{ULA}{uniform linear array}
\acrodef{URA}{uniform rectangular array}
\acrodef{VGA}{variable gain amplifier}
\acrodef{ZF}{zero-forcing}
\acrodef{ZMCG}{zero-mean complex Gaussian}